\documentclass[twocolumn,showpacs,preprintnumbers,amsmath,amssymb]{revtex4-1}
\usepackage{amssymb}
\usepackage{amsmath}
\usepackage{graphicx}
\usepackage{color}
\usepackage{hyperref}

\newcommand{\be}{\begin{equation}}
\newcommand{\ee}{\end{equation}}
\newcommand{\bea}{\begin{eqnarray}}
\newcommand{\eea}{\end{eqnarray}}
\newcommand{\beal}{\begin{aligned}}
\newcommand{\eeal}{\end{aligned}}

\usepackage{color}

\begin{document}

\preprint{DCPT-16/15}


\title{Thermodynamics of Accelerating Black Holes}


%\thanks{}
\author{Michael Appels$^1$}
\email{michael.appels@durham.ac.uk}

\author{Ruth Gregory$^{1,2}$}
\email{R.A.W.Gregory@durham.ac.uk }

\author{David Kubiz\v n\'ak$^2$}
\email{dkubiznak@perimeterinstitute.ca}

\affiliation{$^1$Centre for Particle Theory, Durham University,
South Road, Durham, DH1 3LE, UK\\
$^2$Perimeter Institute, 31 Caroline St, Waterloo, Ontario N2L 2Y5,
Canada}


\date{\today}

\begin{abstract}
We address a long-standing problem of describing the thermodynamics
of a charged accelerating black hole. We derive a standard first law of
black hole thermodynamics, with the usual identification of entropy
proportional to the area of the event horizon -- even though the event
horizon contains a conical singularity.
This result not only extends the applicability of black hole thermodynamics
to realms previously not anticipated, it also opens a possibility for studying
novel properties of an important class of exact radiative solutions of
Einstein equations describing accelerated objects.
We discuss the thermodynamic volume, stability and phase structure
of these black holes.

\end{abstract}


\maketitle

Black holes are possibly the most fascinating objects in our universe. They
provide a practical environment for testing strong gravity, and are also 
incredibly important theoretical tools for exploring Einstein's
General Relativity (GR) and beyond. In spite of their central
importance, the number of exact solutions describing a black hole is
incredibly small; the Kerr-Newman family give us our prototypical 
black hole in four dimensions, these are parametrised simply by
mass, charge and angular momentum. There is however another
exact solution for a black hole, less well known: the {\it C-metric}
\cite{Kinnersley:1970zw, Plebanski:1976gy,Dias:2002mi, Griffiths:2005qp}
that represents an accelerating black hole, a conical deficit angle along one 
polar axis attached to the black hole providing the force driving the acceleration. 
Although this exact 
solution is idealised, the conical singularity pulling the black hole
can be replaced by a finite width cosmic string core \cite{Gregory:1995hd}, 
or a magnetic flux tube \cite{Dowker:1993bt}, and one can imagine that 
something like the C-metric with its
distorted horizon could describe a black hole that has been accelerated by
an interaction with a local cosmological medium.

The C-metric also has applications beyond pure classical GR. It
describes the pair creation of black holes, either in a magnetic or 
electric field \cite{Dowker:1993bt}, and also the splitting of a cosmic string
\cite{Eardley:1995au,Gregory:1995hd}. Its most
important theoretical application was probably in the construction of
the black ring solution in 5D gravity \cite{Emparan:2001wn}. 
The C-metric has also served as a testing ground for the study of
gravitational radiation (see e.g.\ \cite{Podolsky:2003gm}).
Yet in spite of this, it has remained
a somewhat esoteric solution, not fully integrated into the arsenal
of tools for the black hole practitioner. This is partly because the
accelerating black hole is not so well understood
theoretically, a glaring hole being the lack of a prescription for
defining thermodynamics of these solutions.

Black hole thermodynamics
\cite{Bekenstein:1973ur, Bekenstein:1974ax, Hawking:1974sw},
has been an important and
fascinating topic providing key insights into the nature
of black holes and classical gravitational theory, and also
opening a window to quantum gravity. This is especially true
for black holes in anti de Sitter (AdS) space, where thermal
equilibrium is straightforwardly defined  \cite{Hawking:1982dh}
and physical processes correspond via a gauge/gravity duality
to a strongly coupled dual thermal field theory \cite{Witten:1998zw}.
To a large extent the thermodynamic properties of black holes
have been mapped out, with a good understanding of the
role of various asymptotic properties, horizon topologies,
charges, yet to our knowledge there has been no critical 
discussion in the literature of thermodynamics of accelerating
black holes.

In this letter, we seek to address this problem, by presenting a
consistent description of the thermodynamics of an accelerating
black hole. Not only will this bring the C-metric onto a more even
footing with other exact solutions commonly used to model black holes,
but may also allow for an investigation of new and interesting phenomena
in the arena of holography, where it will correspond to a finite temperature 
highly nontrivial system with boundary physics.

\begin{figure}
\includegraphics[scale=0.3,angle=90]{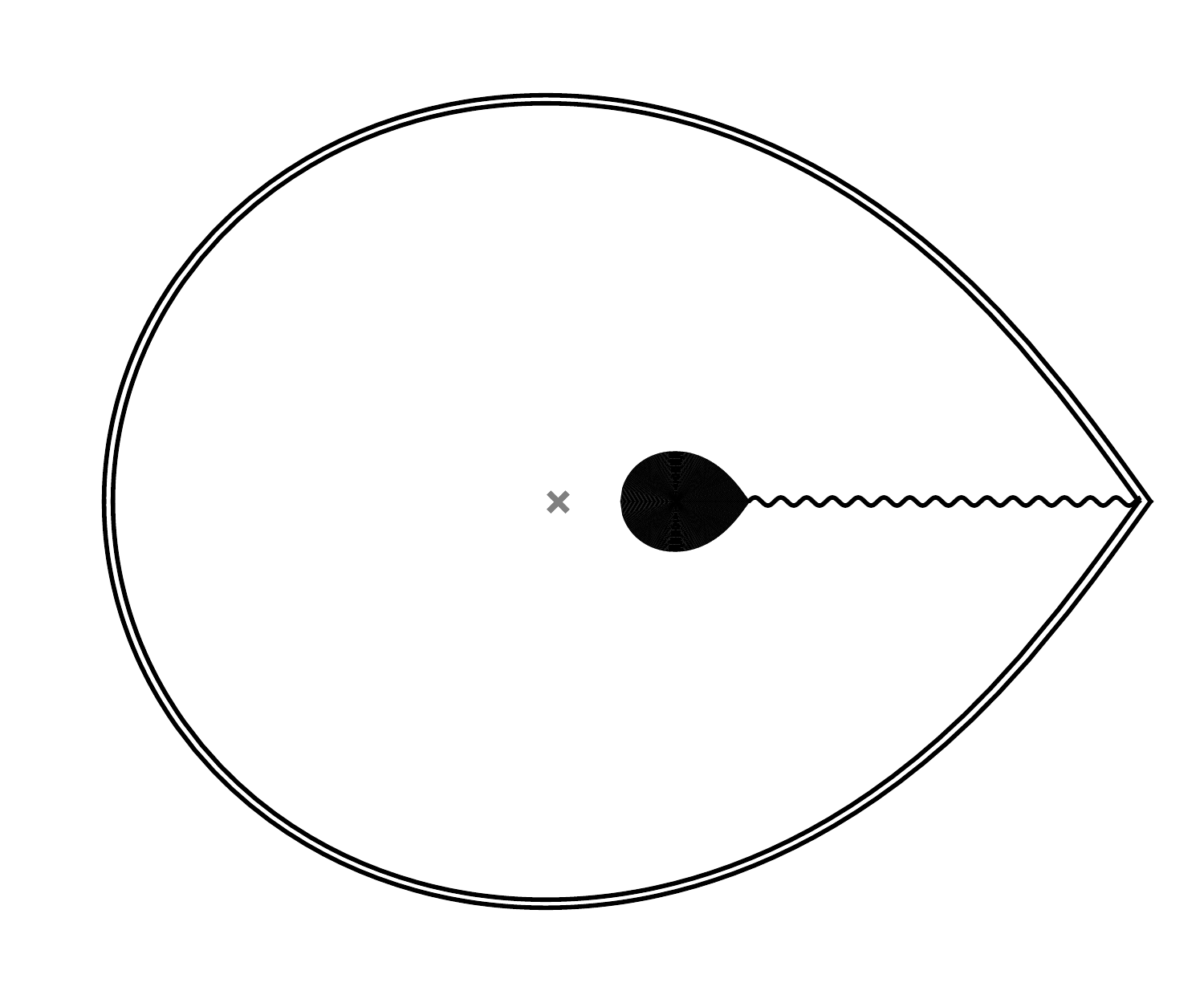}
\caption{A cartoon of the slowly accelerating black hole
in the Poincare disc of AdS: the horizon has a conical shape 
at one pole where the cosmic string (indicated by a wiggly line)
attaches and pulls on the black hole, suspending it
away from the centre of AdS, here shown by a grey cross.
}
\label{fig:accbh}
\end{figure}

One feature of the accelerating black hole is that it
generically has an acceleration horizon due to the fact that
a uniformly accelerating observer asymptotically approaches
the speed of light and hence can never see anything beyond
this asymptotic light cone. The existence of
this second horizon raises the problem of thermodynamic
equilibrium, as one would expect the local temperatures associated
to each horizon to be different. One way around this problem is
to consider a negative cosmological constant that can negate this
effect and `remove' the acceleration horizon. Such a black hole
is said to be {\it slowly accelerating}, and is displaced 
from the center of the negatively curved
space-time at the cost of applying a force in the form
of a cosmic string ending on the black hole horizon
\cite{Podolsky:2002nk}.
Fig.~\ref{fig:accbh} shows a representation of the black hole horizon
with cosmic string.

One advantage of having no additional horizon is that the
temperature of the black hole can now be uniquely defined,
yet the existence of the cosmic string pulling the black hole
off-center means that the black hole is not isolated, and therefore
one should be careful when considering thermodynamic
variations. Furthermore, although the C-metric does not appear
to be time-dependent, an accelerating object carries with it the
notion of some form of time variance, and there is 
non-zero radiation at infinity \cite{Podolsky:2003gm}, which raises the
question: How can a system be in equilibrium if it is accelerating?

Here we will answer these questions, formulating and investigating
the thermodynamics of these slowly accelerating black holes.
We begin by discussing the physics of the accelerating black hole,
explaining the relation between physical quantities and the parameters
in the mathematical solution. By considering the black hole plus
string system as a unit, allowing only physically consistent variations,
we derive a standard First Law of thermodynamics and a Smarr
formula. We will see the accelerating black hole throws up a few new
surprises in terms of the dynamical processes that are allowed.
Finally, we discuss the thermodynamical properties of our
black holes and the existence of a Hawking--Page transition \cite{Hawking:1982dh}.

A charged accelerating AdS black hole is
represented by the metric and gauge potential \cite{Griffiths:2005qp}:
\be
\beal
ds^2&=\frac{1}{\Omega^2}\Bigl[ f(r)dt^2-\frac{dr^2}{f(r)}
-r^2\Bigl(\frac{d\theta^2}{g(\theta)}
+g(\theta)\sin^2\!\theta \frac{d\phi^2}{K^2}\Bigr)\Bigr]\,,\\
F&=dB\,,\quad B=-\frac{e}{r}dt\,,
\eeal
\label{dimlessmtrc}
\ee
where
\be
\beal
f(r)&=(1-A^2r^2)\Bigl(1-\frac{2 m}{r}+\frac{e^2}{r^2}\Bigr)+\frac{r^2}{\ell^2}\;,\\
%g&=(1-x^2)\bigl[1+2mA x+e^2A^2x^2\bigr]\,,
g(\theta)&=1+2mA \cos\theta+e^2A^2\cos^2\theta\,,
\eeal
\label{fandg}
\ee
and the conformal factor
\be
\Omega=1+A r \cos\theta\;
\ee
determines the {\em conformal infinity}, or boundary, of the AdS space-time.
The parameters $m$ and $e$ are related to the black hole mass and electric
charge, $A>0$ is related to the magnitude of acceleration of the black 
hole, and $\ell = \sqrt{-\Lambda/3}$ is the AdS radius.

This particular way of writing the metric gives transparent continuity to
the AdS black hole, and shows how the acceleration distorts the
spherical surfaces (including the horizon) represented by the polar
$\theta, \phi$ angles (see \cite{Hong:2003gx} for a discussion of various coordinates
for the C-metric). Looking at \eqref{fandg}, we see that the acceleration
parameter competes with the cosmological constant ``$r^2/\ell^2$''
term in the Newtonian potential, alternatively, the negative curvature
of AdS space negates the effect of acceleration. It is easy to see from
the form of $f$ that $A<1/\ell$ describes a single black hole suspended in
AdS with the only horizon being that of the black hole
\cite{Podolsky:2002nk}. For $A>1/\ell$ two (oppositely charged)
black holes are present and separated by the acceleration horizon
\cite{Dias:2002mi, Krtous:2005ej}; the case of $A=1/\ell$ is special
and was discussed in \cite{Emparan:1999fd}. We further restrict
$mA<1/2$ so that our angular coordinates correspond to the
usual coordinates on the two-sphere. For a discussion of
general C-metrics in AdS and their holographic implications
see \cite{Hubeny:2009kz}.

The presence of the cosmic string is discovered by looking at the
angular part of the metric and the behaviour of $g(\theta)$ at the
poles, $\theta_+=0$ and $\theta_- = \pi$. Regularity of the metric
at a pole demands
\begin{equation}
K_\pm = g(\theta_\pm)=1\pm 2mA+e^2A^2.
\end{equation}
Clearly, for $mA\neq0$, it is not possible to fix $K$ such that we
have regularity at both poles, and the lack of regularity at an axis
is precisely the definition of a conical singularity. Typically, $K$ is
chosen to regularise one pole, leaving either a conical deficit or
a conical excess along the other pole. Since a conical
excess would be sourced by a negative energy object, we suppose
that our black hole is regular on the North Pole ($\theta=0$), fixing
$K=K_+=1+2mA+e^2A^2$, and then
on the South Pole axis, $\theta=\pi$, there is a conical deficit:
\begin{equation}\label{deficits}
\delta=2\pi\left(1-\frac{g_-}{K_+}\right)
=\frac{8\pi mA}{1+2mA+e^2A^2}\,,
\end{equation}
that corresponds to a cosmic string with tension $\mu=\delta/8\pi$.

To sum up: there are 5 physical parameters in
the C-metric solution: the mass $m$, the charge $e$,
the acceleration $A$, the cosmological constant
represented by $\ell$, and the tension of the cosmic
strings on each axis, encoded by the periodicity of the
angular coordinate. It would seem therefore
that a First Law of thermodynamics could relate variations
in the mass of the black hole to variations in charge,
pressure ($\Lambda$), entropy and acceleration, however
this is not the case.

When considering thermodynamical properties of the black hole,
we must consider physically reasonable variations we can make
on the system that now consists of the black hole plus cosmic
string. Intuitively, if we add mass to the black hole, this will have a
consequence: a more massive object will accelerate more slowly,
thus changing ``$M$'' in the system will also change acceleration.
If the black hole is charged, then changing $Q$ will likewise alter
the acceleration. Given that the cosmic string pulling the black
hole cannot instantaneously change its tension (indeed, if it is
a vortex solution to some field theory model, it cannot change its
tension at all), this means that our thermodynamic variations will
be constrained by the physics of the system.

We start by identifying the relevant thermodynamic quantities.
For the black hole mass we used the method of
conformal completion \cite{Ashtekar:1984zz, Ashtekar:1999jx,
Das:2000cu}. This takes the electric part of the Weyl tensor
projected along the time-like conformal Killing vector, $\partial_t$, and
integrates over a sphere at conformal infinity.
The calculation gives
\be
M=\frac{m}{K}\,,
\ee
thus $m$ gives the mass of the black hole. Note that
unlike the rapidly accelerating black hole, this is a genuine
ADM-style mass, and not a ``re-arrangement of dipoles''
as discussed in \cite{Dutta:2005iy} where a boost mass
was introduced. Similarly,
the electric charge $Q$ and the electrostatic potential
$\Phi$ evaluated on the horizon are
\be
\beal
Q&=\frac{1}{4\pi}\int_{\Omega=0} *F=\frac{e}{K}\,,\\
\Phi&=\frac{e}{r_+}\,.
\eeal
\ee
Meanwhile, we identify the entropy with a quarter of the horizon
area
\be\label{entropy}
S=\frac{\cal A}{4}=\frac{\pi r_+^2}{K(1-A^2r_+^2)}\,,
\ee
and calculate the temperature via the usual Euclidean method to obtain
\be
T=\frac{f'(r_+)}{4\pi} =
\frac{m}{2\pi r_+^2} - \frac{e^2}{2\pi r_+^3}+ \frac{A^2 m}{2\pi}
- \frac{A^2 r_+}{2\pi} + \frac{r_+}{2\pi\ell^2}\,,
\label{temp}
\ee
using $f(r_+)=0$ to collect terms together.
We now identify $P$ with the pressure associated to the
cosmological constant according to
\be
P=-\frac{\Lambda}{8\pi}=\frac{3}{8\pi \ell^2}\,,
\ee
which allows us to rewrite the temperature equation \eqref{temp}
as:
\be
TS= \frac{M}{2} - \frac{\Phi Q}{2} + P\, \frac{4\pi}{3K}
\,\frac{r_+^3}{(1-A^2r_+^2)^2}\,,
\ee
which is nothing other than a Smarr formula \cite{Smarr:1972kt,Kastor:2009wy}
\be
M= 2(TS - PV)+ \Phi Q
\label{smarr}
\ee
provided we identify the black hole thermodynamic volume as
\be\label{V}
V=\frac{\partial M}{\partial P}\Big|_{S,Q}\,=
\frac{4\pi}{3K}\frac{r_+^3}{(1-A^2r_+^2)^2}\,.
\ee

So far, this is a rewriting of a relation for the temperature,
having identified standard thermodynamic variables or
charges for the solution. Now consider the First Law.
Typically, one derives this by observing the change
in horizon radius during a physical process. The horizon
radius is given by a root of $f(r_+)=0$, and thus depends
on $m$, $e$, $A$ and $\ell$. The specific form of this
algebraic root is not vital, what matters is how the mass varies
in terms of the change in horizon area, thermodynamic volume,
and charge.

During this process, any conical deficit cannot
change, as it corresponds to the physical object causing 
acceleration. Thus we must consider a variation of $m$,
$e$ and $A$ that preserves the cosmic string(s), and it turns out that it
is precisely this physical restriction that allows us to derive the First Law.

To obtain the First Law, we typically consider a perturbation of
the equation that determines the location of the event horizon
of the black hole: $f(r_+)=0$. If we allow our parameters to vary,
this will typically result in a perturbation also of $r_+$, hence we
can write
\be
\frac{\partial f}{\partial r_+} \delta r_+ +
\frac{\partial f}{\partial m} \delta m +
\frac{\partial f}{\partial e} \delta e +
\frac{\partial f}{\partial A} \delta A +
\frac{\partial f}{\partial \ell} \delta \ell =0\,,
\ee
where everything is evaluated at $f(r_+,m,e,A, \ell)=0$.
Clearly we can replace $\delta m$, $\delta e$, and
$\delta \ell$ by variations of the thermodynamic parameters
$M, Q$, and $P$, and $\delta r_+$ is expressible in terms
of $\delta S$ and $\delta A$ using \eqref{entropy}. Finally,
we replace $\partial f /\partial r_+ = 4 \pi T$, and use
$f(r_+)=0$ to simplify the terms multiplying $\delta A$ to
obtain:
\be
(1-A^2 r_+^2) (T \delta S + V \delta P) - \delta M + \Phi \delta Q
- \frac{r_+^2 A}{K} \delta A (m-e\Phi) =0\,.
\label{firstint}
\ee
At the moment, it seems as if we have an extra thermodynamic
quantity however, we now use the physical input from the cosmic
string that the conical deficits on each axis must not change.
This means that $\delta K_+(m,e,A)=0$ so that our North Pole
axis remains smooth, and $\delta\mu(m,e,A)=0$ so that our
cosmic string tension is unchanged. These two conditions imply
that $mA$ {\it and} $eA$ are unchanged,
hence $m\delta A = - A \delta m$ and $e \delta A = - A \delta e$.
Replacing $\delta A$ in \eqref{firstint} and re-arranging gives the
First Law:
\be
\delta M = T \delta S + \Phi \delta Q+ V \delta P\,.
\label{first}
\ee

Now that we have unambiguous thermodynamical variables for
our accelerating black hole, we can explore its properties. One
simple consequence is that the black hole satisfies the
{\it Reverse Isoperimetric Inequality}, conjectured for non-accelerating
black holes \cite{Cvetic:2010jb}. The Isoperimetric Inequality
states that the volume enclosed within a given area
is maximised for a spherical surface, this is the reason
soap bubbles are spherical. For black holes, surface area
corresponds to entropy, so from thermodynamical considerations, 
we would expect
that spherical black holes would {\it maximise} entropy otherwise our
black holes would have a different shape. It was precisely this
{\it reverse} inequality that was conjectured and explored in
\cite{Cvetic:2010jb}.

For the slowly accelerating black hole, we therefore want to compare the
volume dependence on $r_+$ to the area dependence via
the isoperimetric ratio
\be
{\cal R}=\Bigl(\frac{3V}{\omega_{2}}\Bigr)^\frac{1}{3}
\Bigl(\frac{\omega_{2}}{{\cal A}}\Bigr)^\frac{1}{2}\,,
\ee
where $V$ is the thermodynamic volume, ${\cal A}$ is the horizon area,
and  $\omega_2=4\pi/K$ is the area of a unit `sphere'.
Using the above formulae for $V$ and ${\cal A}$, we find
\be
{\cal R}=\frac{1}{(1-A^2r_+^2)^{1/6}}\geq 1\,.
\ee
Thus these slowly accelerating black holes do indeed satisfy the
reverse isoperimetric inequality.

Another fascinating aspect of black holes in AdS is
that, unlike asymptotically flat black holes, they are not always
thermodynamically unstable. A Schwarzschild black hole loses mass
through Hawking radiation, becoming hotter and eventually evaporating
away. In AdS however, black holes larger than of order the AdS radius
instead become cooler as they lose mass, and indeed are thermodynamically
stable as demonstrated by the form of their Gibbs free energy.

\begin{figure}
\begin{center}
\rotatebox{0}{
\includegraphics[width=0.35\textwidth]{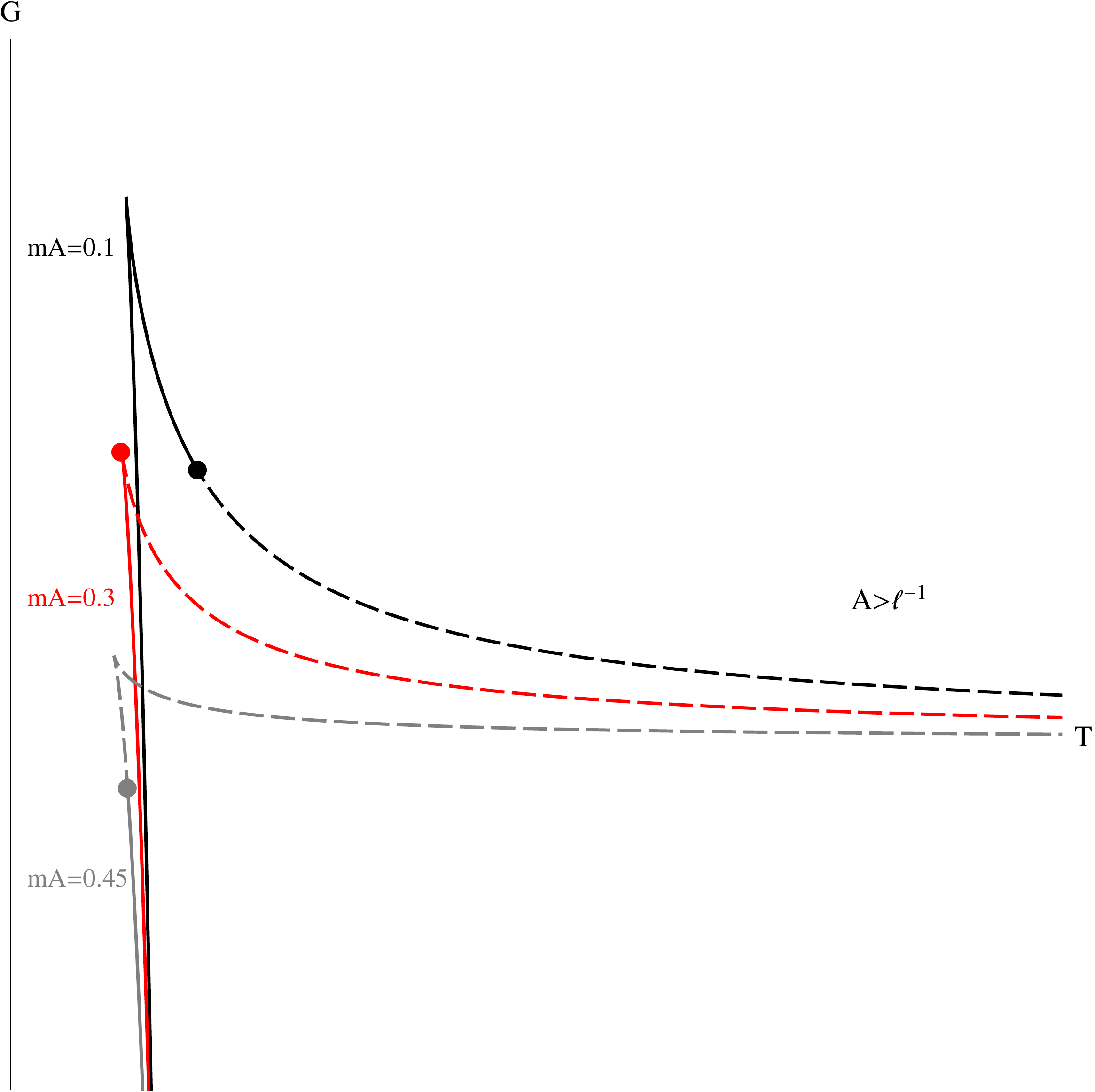}
}
\caption{Plots of the Gibbs function as a function of $T$ at fixed $P$.
We explore varying the cosmic string tension, represented
by $mA$. The solid lines represent the slowly accelerating black holes. 
The dot is the point $A\ell=1$, although we have continued the plot 
for $A>1/\ell$ shown by the dashed lines.} 
\label{fig:gibbs}
\end{center}
\end{figure}

Focussing on the uncharged slowly accelerating black hole,
and constructing the associated Gibbs free energy,
\be
G=G(P,T)=M-TS\,,
\ee
we display the behavior of $G=G(P,T)$ in Fig.~\ref{fig:gibbs}, showing
how it depends on the tension of the cosmic string encoded by $mA$. 
The behaviour of $G$ is reminiscent of the Hawking--Page phase 
transition \cite{Hawking:1982dh},
however, in this spacetime we have a conical singularity (with fixed
deficit angle) that extends to the AdS boundary. It is therefore not
possible to have a phase transition between a pure radiation AdS
spacetime to the accelerating black hole. We also emphasise that
different points on the curve correspond not only to different size
but also differently accelerated black holes. As expected, the black holes on the
upper branch of the curve have negative specific heat, and those
on the lower branch, positive specific heat.

Slowly accelerating black holes are therefore very similar to their non-accelerating
cousins from a thermodynamical perspective. One interesting difference lies
in the constraint coming from the cosmic string suspending the black hole.
By taking this string to be an approximation to a physical object, we
conclude that it cannot change tension, and this translates into
constraints on the allowed variations of the black hole. Both the
mass and charge can vary, but they must vary in the same way
keeping $mA$ and $eA$ constant. In the absence of charge, this makes
perfect sense from Newton's First Law: if an object gets heavier,
but is subject to the same force, then it will accelerate more slowly.
However, the behaviour of the charged accelerating black hole
is far more interesting; it would seem that we cannot throw an
uncharged mass into the black hole. Once an accelerating black hole
has charge, the Maxwell field no longer vanishes on the boundary
$Ar\cos\theta =-1$:
\be
F = d \left [ eA\cos\theta dt \right]
= e A \sin\theta \, dt \wedge d\theta\,,
\ee
thus if the acceleration of the black hole were to change without changing
its charge, the electric field on the boundary would also have to change.

It is worth noting that this situation is remarkably similar to the
thermodynamics of (charged) Taub-NUT-AdS spacetimes studied
in, e.g., \cite{Chamblin:1998pz,Johnson:2014pwa}.
There, AdS spacetimes with a NUT charge were considered,
and a constraint on the periodicity of Euclidean time, similar to the imposition of
the constant deficit in our accelerating black hole, has to be 
imposed in order that a Misner string is not observable in the 
spacetime. This is then used to confirm the usual `First law'. 
Even more remarkably, again similar to our situation, in the 
presence of charge the regularity of the charged NUT-AdS 
solution requires two conditions, one imposed on
the temperature, the other on the charge, so that the first law can
hold \cite{Johnson:2014pwa}. However, there is one crucial
difference: in the Taub-NUT case the entropy is not given by the
Bekenstein--Hawking area law, but rather, is derived from demanding
the First Law. Further similarities and differences between these two
classes of geometries will be studied elsewhere.

Finally, it is interesting to consider possible extensions of these
results. Here, we fixed the conical deficits, motivated by the physical
assumption that they were representative of a physical source, such 
as a cosmic string. However, in principle, one could vary tension -- 
metrics with multiple accelerating black holes and tensions are 
known \cite{Dowker:2001dg}. Generalising our results
to this more interesting and complex case is underway. It is also worth
remarking that our discussion here is restricted to four dimensions, as a
C-metric in general dimensions has so far proven to be elusive. However,
if one considers accelerating black branes (rather than holes) then 
presumably the methods here could be applied to a wider family of
black branes in arbitrary dimensions.

\begin{acknowledgments}
We would like to thank Claude Warnick for sharing unpublished work
on an alternative approach to thermodynamics of the C-metric.
We would also like to thank Pavel Krtous, Rob Myers, Simon Ross
and Jennie Traschen for useful conversations.
MA is supported by an STFC studentship.
RG is supported in part by STFC (Consolidated Grant ST/L000407/1),
in part by the Wolfson Foundation and Royal Society, and also
by the Perimeter Institute for Theoretical Physics. DK is also
supported in part by Perimeter, and by the NSERC.
Research at Perimeter Institute is supported by the Government of
Canada through Industry Canada and by the Province of Ontario through the
Ministry of Research and Innovation.

\end{acknowledgments}

\providecommand{\href}[2]{#2}\begingroup\raggedright\endgroup

\end{document}